\documentclass[epjST_arxiv]{svjour_arxiv}
\usepackage{graphicx}
\usepackage{epstopdf}
\usepackage{amsmath}
\usepackage{amssymb}
\usepackage{stackengine,scalerel}
\def\lambdabar{\ThisStyle{\ensurestackMath{\stackon[-2.4\LMpt]{%
  \SavedStyle\lambda}{\kern-.5pt\kern\LMpt\rule{1\LMex}{.25pt+.15\LMpt}}}}}
\def\Foerster{F\"orster\xspace}

\newcommand{\blst}{\begin{lstlisting}}
\newcommand{\est}{\end{lstlisting}}

\newcommand{\bit}{ \begin{itemize}{}  }
\newcommand{\eit}{\end{itemize}}

\newcommand{\be}{\begin{enumerate} \itemsep -4pt  }
\newcommand{\ee}{\end{enumerate}}

\newcommand{\bi}{\begin{itemize}   } 
\newcommand{\ei}{\end{itemize}}

\newcommand{\bma}{\begin{math}} 
\newcommand{\ema}{\end{math}}

\newcommand{\bc}{\begin{columns}} 
\newcommand{\ec}{\end{columns}}

\newcommand{\bbl}{\begin{block}} 
\newcommand{\ebl}{\end{block}}

\newcommand{\bflsh}{\begin{flashcard}}
\newcommand{\eflsh}{\end{flashcard}}

\newcommand{\bfl}[2]{\begin{flashcard}{#1} {#2} \eflsh}

\newcommand{\beq}{\begin{equation} \vspace{-0em}} 
\newcommand{\eeq}{\vspace{-0.em} \end{equation}}

\newcommand{\beqs}{\begin{equation*}}
\newcommand{\eeqs}{\end{equation*}}

\def\beqa{\begin{eqnarray}}
\def\eeqa{\end{eqnarray}}

\newcommand{\beal}{\begin{align}}
\newcommand{\eeal}{\end{align}}

\newcommand{\bvr}{\begin{verbatim}}
\newcommand{\evr}{\end{verbatim}}

\newcommand{\om}{\ensuremath{\omega}\xspace}

\newcommand{\dr}{^{\dagger} }

\newcommand{\figref}[1]{FIG.~\ref{#1}}
 
\newcommand{\bra}[1]{\left\langle#1\right|}  								
\newcommand{\ket}[1]{\ensuremath{\left|#1\right\rangle}}  							 	
\newcommand{\braketop}[3]{\left\langle#1\middle|#2\middle|#3\right\rangle}	
\newcommand{\ketbra}[2]{\left|#1\rangle\langle#2\right|}  			

\renewcommand{\vec}[1]{\boldsymbol{#1}}                             

\newcommand{\integralb}[3]{\int\limits_{#1}^{#2} \! \mathrm{d} #3\,} 	
\newcommand{\integral}[1]{\int \! \mathrm{d} #1\,}                    
\newcommand{\integralf}[2]{\int \! \frac{\mathrm{d} #1}{#2}\,}        

\def\bx{{\bf x}}

\def\bR{{\bf R}}

\def\d{{\rm d}}

\def\nn{\nonumber}

\def\a{\ensuremath{\alpha}\xspace}

\def\b{\ensuremath{\beta}\xspace}
\def\g{\ensuremath{\gamma}\xspace}
\def\d{\ensuremath{\delta}\xspace}
\def\D{\ensuremath{\Delta}\xspace}
\def\O{\ensuremath{\Omega}\xspace}

\newcommand{\rs}{\rm \scriptscriptstyle}

\def\L{\ensuremath{\mathcal{L}}\xspace}

\def\Hnon{{H}_{0}\xspace} 
\def\rhoOp{{\rho}\xspace}
\def\rhoSS{\ensuremath{\rho_{ss}\xspace}}
\def\rhoSE{\ensuremath{\rho_{se}\xspace}}
\def\rhoES{\ensuremath{\rho_{es}\xspace}}

\def\gSS{\ensuremath{\gamma_{s}^{\rs deph}\xspace}}
\def\gS{\ensuremath{\gamma_{s}\xspace}}
\def\sSS{\ensuremath{P\xspace}}
 
\def\TM{\ensuremath{T}}
\def\aHbarC{\ensuremath{\frac{\a}{\hbar c}}}
\def\chiBetaSq{\ensuremath{\chi^{s\b}(\om,q)}}
\def\psiWf{\ensuremath{\varphi}}
\def\psiWfE{\ensuremath{\varphi^e\xspace}}
\def\psiCond{\ensuremath{\Psi}}
\def\psiOp{\ensuremath{\psi}}
\def\psiOpE{\ensuremath{\psi_e}}
\def\psiOpP{\ensuremath{\psi_p}}
\def\psiOpS{\ensuremath{\psi_s}}
\def\FOp{\ensuremath{F}}

\def\pSt{\ensuremath{p}}
\def\rSt{\ensuremath{s}}
\def\DeltaDeriv{\ensuremath{\delta}}

\def\z{z}
\def\r{r}
\def\q{k}
\def\gamP{\ensuremath{\gamma}}

\def\Veff{\ensuremath{V^{\rs eff}}}
\def\VthreeD{\ensuremath{V^{\rs 3D}}}
\def\alphaInd{\mu} 
\def\aInd{\mu}

\def\fullG{\ensuremath{\mathcal{G}}}
\def\kB{\ensuremath{k_{\rs B}}}
\def\chiB{\ensuremath{\bar{\chi}}}
\def\vDelta{\ensuremath{v}}
\def\ESt{\ensuremath{e}}
\def\HCoh{\ensuremath{\mathcal{H}}\xspace}
\def\rhoOne{\ensuremath{\rho^{(1)}}\xspace}
\def\Schroedinger{Schr\"odinger\xspace}
\def\rhoOpN{\ensuremath{\rho^{(N)}}\xspace}
\def\HInt{\ensuremath{H_{\rs rr}}\xspace}
\def\ba{\begin{eqnarray}}
\def\ea{\end{eqnarray}}
\def\beq{\begin{equation}}
\def\eeq{\end{equation}}

\def\etal{\textit{et al.}\xspace} 
\usepackage{xspace} 
\usepackage{color}
\makeatletter
\g@addto@macro\@floatboxreset\centering
\makeatother

\begin{document}
\title{Few-body quantum physics with strongly interacting Rydberg polaritons}
\author{Przemyslaw Bienias\inst{1}\fnmsep\thanks{\email{przemek@itp3.uni-stuttgart.de}} }
\institute{
Institute for Theoretical Physics III and Center for Integrated Quantum Science and Technology, Universit\"{a}t Stuttgart, Pfaffenwaldring 57, 70569 Stuttgart, Germany}
\abstract{
We present an extension of our recent paper [Bienias \etal, Phys. Rev. A \textbf{90}, 053804 (2014)] in which we demonstrated the scattering properties and bound-state structure of two Rydberg polaritons, as well as the derivation of the effective low-energy many-body Hamiltonian.
Here, we derive a microscopic Hamiltonian describing the propagation of Rydberg slow light polaritons in one dimension. 
We describe possible decoherence processes within a Master equation approach, and
derive equations of motion in a \Schroedinger picture by using an effective non-Hermitian Hamiltonian.
We illustrate diagrammatic methods on two examples: 
First, we show the solution for a single polariton in an external potential by exact summation of Feynman diagrams.
Secondly, we solve the two body problem in a weakly interacting regime exactly. 
} 
\maketitle
\setcounter{tocdepth}{4}
\setcounter{secnumdepth}{4}

\section{Introduction}

The photon interacts with its environment much weaker than other quanta (e.g. electron spin, superconducting current), and is therefore an excellent carrier of information. 
A long-standing goal is the realization of strong interaction between individual photons which 
may lead to ultralow-power all-optical signal processing~\cite{Hu2008a,Miller2010}, quantum information processing and communication~\cite{Milburn1989,Kimble2008}
as well as other applications based on non-classical states of light~\cite{Muthukrishnan2004,Giovannetti2011}.

A number of promising platforms to engineer suitable interactions between photons are being developed~\cite{Chang2014}.
First ideas were based on the Kerr nonlinearity of conventional materials, 
which unfortunately leads to extremely weak effects for single photons, even for highly nonlinear fibres~\cite{Matsuda2009}.
In the microwave domain, a significant progress was done using high-quality-factor (high-Q) cavity quantum electrodynamics where a single confined electromagnetic mode is coupled to an atomic system~\cite{Fushman2008,Rauschenbeutel1999,Devoret2013}, or to a superconducting circuit acting as an `artificial atom'~\cite{Kirchmair2013,Haroche2006}.
In the optical domain, one approach is to map photons onto the collective states of an atomic ensemble~\cite{Fleischhauer2005a,Hammerer2010}, which enabled the observation of electromagnetically-induced-transparency cross-coupling nonlinearities~\cite{Bajcsy2009} and similar effects~\cite{Venkataraman2011}. 
However, in order to achieve single photon effects~\cite{Tanji-Suzuki2011,Chen2013a} cavities are necessary.

A promising approach, that does not require optical resonators, capitalizes on strong atom-atom interactions in the metastable Rydberg state  $\ket{\rSt}$ of an EIT scheme \cite{Lukin2001,Friedler2005}. 
These strong, tunable and long-range interactions~\cite{Saffman2010bb} enabled a number of applications for quantum computing~\cite{Wilk2010,Isenhower2010a,Weimer2010aa,Jau2016} and simulations~\cite{Schauss2012,Schauss2015,Weber2015a,Glaetzle2015,VanBijnen2015}, in which quantum information is encoded in the atomic degrees of freedom. 
Using Rydberg-Rydberg interactions to enhance nonlinearities between photons was first proposed by Friedler \etal in 2005~\cite{Friedler2005}, then extended to the studies of many-body correlations~\cite{Sevincli2011,Petrosyan2011}, and 
photonic quantum gates~\cite{Gorshkov2011a}.
On experimental side~\cite{Mohapatra2007,Pritchard2010}, this approach enabled a variety of applications such as deterministic single photon source~\cite{Dudin2012}, observation of the quantum phenomena on a few photon level~\cite{Parigi2012}, atom-photon entanglement generation~\cite{Li2013}, a single photon switch~\cite{Baur2014}, a transistor~\cite{Gorniaczyk2014,Tiarks2014}, a single photon absorber~\cite{Tresp2016}, and a phase gate~\cite{Tiarks2015}.
Moreover, in the regime of strong interaction between copropagating photons the medium transparent only to single photons~\cite{Peyronel2012} as well as the bound states of photons~\cite{Firstenberg2013} were demonstrated.

First Rydberg-EIT experiments mostly exploited interaction between $s$-states with the same principal quantum  number, where the angular dependence of the interaction is very weak.
In recent experiments, Rydberg $s$- and \pSt-states~\cite{Maxwell2013}, or two different $s$-states~\cite{Gorniaczyk2014,Tiarks2014,Gunter2013} are simultaneously prepared. 
The interaction between different Rydberg states enables novel entanglement schemes~\cite{Cano2014} and additional flexibility in manipulation of few-photon light fields~\cite{Li2015a,Liu2015}.
Moreover, 
the angular dependence of the interaction between $d$-state polaritons can lead to the interaction-induced dipolar dephasing of polariton pairs~\cite{Tresp2015}.

The additional  tunability of the interactions was investigated in setups close to \Foerster resonance,  where the interaction changes its character from van-der-Waals to dipolar.
In the regime of zero electric-field, an enhancement of the single-photon-transistor gain was shown~\cite{Tiarks2014}, 
while in the experiments on Rydberg atom imaging~\cite{Olmos2011,Gunter2012}, performed in the regime of many photons, an increase in Rydberg excitation hopping was observed~\cite{Gunter2013}.
Furthermore, the Stark-tuned \Foerster resonances were used to further improve the efficiency of the transistor and to study its coherent properties~\cite{Gorniaczyk2015}. 

The extensions of the two-qubit photonic gate~\cite{Gorshkov2011a}, 
employing spatial separation of photons~\cite{He2014a} and performing quantum operations on the stored photons~\cite{Khazali2015,Paredes-Barato2014a},  were proposed. 
In the one-dimensional free-space setups realized so far, the 
achievable optical depth per blockade radius is limited by the interaction between ground state and Rydberg atoms~\cite{Gaj2014a}. 
This, in turn, sets constraints on the available amount of the dispersive interaction per photon for the quantum information applications.
To circumvent this limitation, an optical cavity can be employed to enhance the interaction per photon life-time~\cite{Stanojevic2013,Grankin2014} and to construct high fidelity phase gate~\cite{Das2016}.

Recently, it was shown that a Rydberg-EIT setup can give rise to new few- and many-body states of light.
Two photons can form shallow~\cite{Firstenberg2013} and deep bound states~\cite{Bienias2014}, which can be imagined as photons trapped by a Rydberg interaction in a deep nearly-square well.
Pair of photons can also interact via an effective Coulomb potential, leading to the hydrogen-like diatomic molecule, separated by a finite bond length~\cite{Maghrebi2015c}.
Finally, the formation of a Wigner crystal of individual photons was predicted~\cite{Otterbach2013,Moos2015a}.

It is also worth mentioning the progress towards many-body theory of strongly interacting Rydberg polaritons~\cite{Bienias2016,Jachymski2016,Gullans2016}.
In the dissipative regime, the dynamics of quantized light was analyzed in~\cite{Gorshkov2013}. 
In the dispersive regime, the derivation of low-energy Hamiltonian in the dilute regime was presented in~\cite{Bienias2014}, whereas in~\cite{Otterbach2013} for higher densities when
 the interaction is dominated by the purely repulsive part of the van der Waals interaction.
Moreover, a recently developed general input-output formalism to describe the dynamics of propagating strongly interacting photons in 1D~\cite{Caneva2015,Shi2015} can be applied to the Rydberg-polariton systems as well.

Most of the initial Rydberg-EIT research investigated effectively one-dimensional systems.
A new promising direction are extensions to higher dimensions. 
For example, Rydberg-dressed photons in near-degenerate optical cavities can behave as interacting, massive, harmonically trapped, two-dimensional particles in a synthetic magnetic field~\cite{Sommer2015}.
The experimental progress~\cite{Ningyuan2016} makes Rydberg-cavity polaritons a promising platform for creating photonic quantum materials and topological states of light~\cite{Maghrebi2015}.

In our previous paper~\cite{Bienias2014}  we used diagrammatic methods to analyze scattering properties and the bound-state structure of two Rydberg polaritons in one dimension.
This framework enabled us to analytically derive the effective interaction potential between two polaritons and to determine a regime with purely repulsive interactions. In the regime of attractive interaction we identified multiple bound states of two polaritons and studied their dispersion relation.  Finally, the derivation of the low-energy scattering length enabled us to microscopically derive the many-body theory for Rydberg polaritons in the dilute regime. 

In the  present paper, we extend our studies from Ref.~\cite{Bienias2014} by starting from the derivation of a microscopic Hamiltonian describing the polariton propagation in one-dimensional free-space. We analyze decoherence processes using a Master equation approach, and show for which processes the evolution can be described using the \Schroedinger equation with an effective non-Hermitian  Hamiltonian. Using our approach, we present a straightforward derivation of equations of motion in \Schroedinger picture for the example of two polaritons. Next, we apply diagrammatic methods to the setup consisting of a single polariton propagating in an external potential, for which we show the exact solution by a summation of all Feynman diagrams. Finally, we present the exact solution of a two-body problem  in a weakly interacting regime. The last result facilitates better understanding of losses from the dark state polaritons to the bright polaritons, discussed in Ref~\cite{Bienias2014}.

\section{Microscopic Hamiltonian derivation}
In this section, we derive a microscopic Hamiltonian describing the
propagation of a weak probe light pulse through an 
 atomic medium under the EIT condition. 

We start with the description of single photons propagating along the $z$-axis in a free space.
In the following, the relevant modes
have only small deviations from the carrier probe frequency $\om_c$ and momentum $\hbar k_c=\hbar\om_c/c$. 
Moreover, we will study the experimentally relevant one-dimensional setup.
The light field distribution $u_k$, characterized by a single transverse mode $u_\perp$,
has the form
\begin{equation}
   u_{k}({\bf x}) =\frac{ e^{i z (k_c +k )}}{\sqrt{L_q}} u_{\perp}({\bf R}),
\end{equation}
with $L_q$ being the quantization length.
For each longitudinal mode $k$,
we introduce the creation operator $a^{\dag}_{k}$. 
Then, the electric field operator reduces to
\begin{equation}
{\bf E}({\bf x}) = \sqrt{\frac{\hbar \omega_{c}}{2\epsilon_0}} \sum_{k} \left[ \vec{\varepsilon}_{k} u_{k}({\bf x}) a^{\dag}_{k} +
\vec{\varepsilon}_{k}^{*} u^{*}_{k}({\bf x}) a_{k}\right],
\end{equation}
with the polarization $\vec{\varepsilon}_{k}$. 
Here, each mode is characterized by a shift in energy from the leading frequency $\omega_c$.
This gives rise to the Hamiltonian in the rotating frame
\begin{equation}
H_{\rs ph} = \sum_{k} \hbar k c\; a^{\dag}_{k} a_{k}
\label{photonHam}
\end{equation}
with $\hbar k c \ll \hbar \omega_{c}$.

Next, we study the interaction of a single photon with the atoms in the medium. For each atom, there are three relevant states within the EIT setup: 
\begin{figure}[htp]
\includegraphics[width= 1\columnwidth]{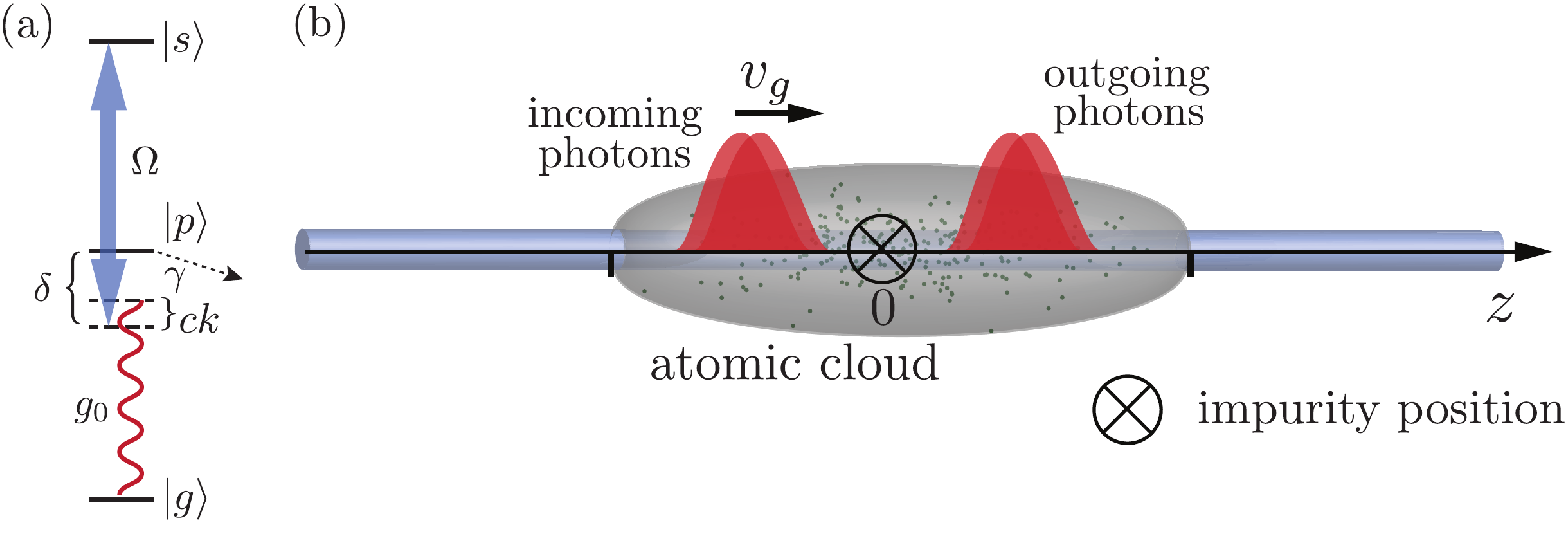}
\caption{
(a) Setup for the electromagnetically induced transparency: the probe field couples the atomic ground state $|g\rangle$ to the $p$-level $|p\rangle$
with the single-particle coupling strength $g_{0}$, while a strong coupling laser drives the transition between the $p$-level and the Rydberg
state $|s\rangle$ with Rabi frequency $\Omega$ and detuning $\delta$.
Furthermore, $2  \gamP$  denotes the decay rate from the $p$-level. The single-particle coupling $g_0$ is related to the collective coupling $g = \sqrt{n} g_0$ with $n$ the particle density. 
(b) Single photons propagate through the atomic medium with the reduced group velocity $v_g\ll c$.
In this paper we will be interested in two scenarios. 
The first, in which a single polariton propagates in an external potential generated by, for example, a stored Rydberg excitation (impurity). 
And second, a scenario in which two polaritons copropagate and interact only with each other.
}
\label{fig1}
\end{figure}
The ground state $|g\rangle$, an intermediate state $|\pSt\rangle$, and finally the Rydberg state $|\rSt\rangle$, see Fig. \ref{fig1}(a).
Within the rotating frame and using the rotating wave approximation, the strong coupling between the intermediate 
state $|\pSt\rangle$ with detuning $\delta$ and the Rydberg state $|\rSt\rangle$ with Rabi frequency 2$\Omega$ gives rise 
to dressed states
\begin{equation}
\begin{array}{lcr}
 |+\rangle = \alpha |\pSt \rangle + \beta |\rSt\rangle,   &\qquad &\Delta_{+} =\left(\DeltaDeriv+ \sqrt{\DeltaDeriv^2 + 4\Omega^2}\right)/2,\\
 |-\rangle = \beta^{*} |\pSt \rangle -  \alpha^{*} |\rSt\rangle, &  \qquad&  \Delta_{-} =  \left(\DeltaDeriv- \sqrt{\DeltaDeriv^2 + 4\Omega^2}\right)/2,
 \end{array}
\end{equation}
with energies $\hbar\Delta_{\pm}$. 
Note, that the spontaneous emission from the intermediate state with decay rate 2\g can be incorporated
by replacing $\delta$ with a complex detuning $\Delta = \delta- i \gamP$, for details see Section \ref{masterConnection}. 
The probe light modes couple the states $|g\rangle$ with $|\pSt\rangle$. 
For simplicity, we restrict the analysis to a situation
where only a single polarization couples  matter with light, with the dipole moment 
$d = \langle g| {\bf d} \cdot \vec{ \varepsilon}|\pSt\rangle$. 
Then, the Hamiltonian, describing the interaction between
the atoms and the light modes reduces to
\begin{equation}
  H_{\rs lm}=   \hbar g \sum_{i}  \left[ \psiOpE({\bf x}_{i}) \: |\pSt\rangle\langle g|_{i} +  \psiOpE^{\dag}({\bf x}_{i})  \: |g\rangle \langle \pSt|_{i} \right]
\end{equation}
with $g = \sqrt{\frac{\omega_{c}}{2\hbar\epsilon_0}} \: d$. 
In addition, we have introduced the field operator for the electric field
\begin{equation}
 \psiOpE^{\dag}({\bf x}) = \sum_{k}  u_{k}({\bf x}) a_{k}^{\dag}.
\end{equation}
In the following, in the continuum limit, we describe the atoms by a field operator   $\bar{\psiCond}$, with the internal structure of the atoms properly accounted for 
by a spinor degree of freedom of $  \bar{\psiCond}$, i.e., 
\begin{equation}
  \bar{\psiCond}({\bf x}) = \left(\begin{array}{c} \psiCond_{g}({\bf x})\\ \psiCond_{+}({\bf x})\\ \psiCond_{-}({\bf x})\end{array}\right).
\end{equation}
This operator can either be a fermionic or a bosonic field operator, depending on the statistics of the atoms. 
Next, we define two new operators, describing
a transition from the ground state $|g\rangle$ into an excited state $|\pm\rangle$,
\begin{eqnarray}
   b^{\dag}({\bf x}) &= &\frac{1}{\sqrt{n({\bf x})}} \: \bar{\psiCond}^{\dag}({\bf x}) S^{+}\bar{\psiCond}({\bf x}),\\
   c^{\dag}({\bf x}) & =&\frac{1}{\sqrt{n({\bf x})}} \: \bar{\psiCond}^{\dag}({\bf x}) T^{+}\bar{\psiCond}({\bf x}),
\end{eqnarray}
with the spinor operator $S^{+}=|+\rangle \langle g|$ and  $T^{+}=|-\rangle \langle g|$. 
In addition, $n({\bf x})$ denotes the atomic density. Then, these operators always satisfy
the bosonic commutation relation,  for example: 
\begin{equation}
  \left[  b({\bf x}), b^{\dag}({\bf y})\right] = \frac{\psiCond^{\dag}_{g}\psiCond_{g} - \psiCond^{\dag}_{+}\psiCond_{+} }{n({\bf x})} \:  \delta({\bf x}-{\bf y}) \simeq \delta({\bf x}-{\bf y}). 
\end{equation}
Here, we have used that fact, that the atomic density is much higher than the photon (polariton) density and, therefore, almost
all atoms are in the ground state, i.e., $\psiCond^{\dag}_{g}\psiCond_{g} \simeq n({\bf x}) \gg \psiCond^{\dag}_{+}\psiCond_{+}$. It immediately follows,
that $b^{\dag}$ and $c^{\dag}$ are bosonic field operators, and the Hamiltonian describing the light field and the interaction with
the atoms reduces to a quadratic Hamiltonian for three coupled bosonic fields,
\begin{equation}
 H = 
 \hbar
 \integral{\bf x} \left(
     \begin{array}{c}  
      	    \psiOpE^{\dag}\\ 
	    b^{\dag}\\ 
	    c^{\dag}
     \end{array}\right)
     \left(\begin{array}{ccc}
        -ic\partial_z
         & g \sqrt{n} \alpha &   g \sqrt{n} \beta\\
         g \sqrt{n} \alpha &    \Delta_{+} & 0\\
        g \sqrt{n} \beta  & 0 &  \Delta_{-}
    \end{array} \right)
        \left(
     \begin{array}{c}  
      	    \psiOpE\\ 
	    b\\ 
	    c
     \end{array}\right),
     \label{quadraticHamiltonian}
\end{equation}
where we have Fourier-transformed the photonic part.

The Hamiltonian in Eq.~(\ref{quadraticHamiltonian}) may be written in a more convenient way by introducing
the fields  $\psiOpP^{\dag}({\bf x}) = -\beta b^{\dag}({\bf x}) + \alpha c^{\dag}({\bf x})$ and $\psiOpS^{\dag}({\bf x}) =  \alpha b^{\dag}({\bf x}) + \beta c^{\dag}({\bf x})$.  These operators describe bosonic fields for the creation of excitation in $|\pSt\rangle$-state and $|\rSt\rangle$-state, respectively. 
Then, the Hamiltonian reduces to
\begin{equation}
 H = \hbar \integral{\bx}\left(
     \begin{array}{c}
      	    \psi_{e}\\
	    \psi_{p}\\
	    \psi_{s}
     \end{array}\right)^{\dag}
     \left(\begin{array}{ccc}
        -ic\partial_z& g    &  0 \\
         g   &   \DeltaDeriv &  \Omega\\
        0  &  \Omega & 0
    \end{array} \right)
        \left(
     \begin{array}{c}
      	    \psi_{e}\\
	    \psi_{p}\\
	    \psi_{s}
     \end{array}\right).
     \label{quadraticHamiltonian-3D}
\end{equation}
Note, that our derivation can be straightforwardly generalized to the light fields confined in a cavity.

Next, we integrate out the transverse mode $u_\perp$  in order to arrive at a one-dimensional theory.
Assuming a homogeneous particle distribution along the longitudinal mode, the light field couples to the following matter mode
\begin{equation}
   \psi_{p}^{\dag}({\bf x}) =   \sqrt{\frac{n({\bf R})}{\bar{n}}}u_{\perp}({\bf R}) \psi_{p}^{\dag}(z)
\end{equation}
with the effective particle density
\begin{equation}
  \bar{n} = \integral{\bR}  n({\bf R}) |u_{\perp}({\bf R})|^2 .
\end{equation}
Analogously, we can define the one-dimensional field operator $\psi_{s}(z)$ accounting for the Rydberg state. Then, 
the operators $\psi_{e}({z})$, $\psi_{p}({ z})$, $\psi_{s}({ z})$ describe a  one-dimensional field theory with the Hamiltonian
\begin{equation}
 \HCoh = \hbar \integral{z} \left(
     \begin{array}{c}  
      	    \psi_{e}^{\dag}\\ 
	    \psi_{p}^{\dag}\\ 
	    \psi_{s}^{\dag}
     \end{array}\right)
     \left(\begin{array}{ccc}
   - i  c \partial_{z}& g    &  0 \\
         g   &    \DeltaDeriv &  \Omega\\
        0  &  \Omega & 0
    \end{array} \right)
        \left(
     \begin{array}{c}  
      	    \psi_{e}\\ 
	    \psi_{p}\\ 
	    \psi_{s}
     \end{array}\right).
     \label{quadraticHamiltonianCoh}
\end{equation}

\section{Decoherence description within Master equation approach \label{masterConnection}}
In this section, we analyze the decoherence of Rydberg polaritons within the formalism developed in the previous section.
The source of the decoherence can be, e.g., spontaneous emission from excited states, motional dephasing or dephasing caused by the interactions between ground and Rydberg states.
In order to understand the impact of these processes on polaritons, we study the system evolution using  the Master equation. In the case of a Markovian evolution, it can be written in the Lindblad form
\ba
\dot{\rho}=-\frac{i}{\hbar}[\HCoh,\rho]+\sum_i\mathcal{L}_i(\rho),
\label{master}
\ea
where $\L_i$ are Liouvillians describing different  incoherent processes, \HCoh describes the coherent evolution and $\rho$ is a density matrix 
\ba
\rho=\sum_{n=0}^N \rho^{(n)},
\label{density}
\ea
where $\rho^{(n)}$ contains $n$ excitations (atomic or photonic).
Moreover,  we neglect correlations between the $N+1$ terms in \eqref{density}.
Note that we truncated the Hilbert space by introducing the maximal number of excitations  $N$ present in the system. 
Such a cut-off is justified for most of the experiments investigating quantum phenomena on a few-photon level with Rydberg-polaritons~\cite{Tiarks2014,Gorniaczyk2014,Peyronel2012,Firstenberg2013,Gorniaczyk2015}. In these experiments, a low intensity laser field is used as a photon source and, thus, the probability of having $N$ excitations in the system is much higher than the probability of having $N+1$ excitations.

As an example, let us consider the case of a single incoming photon, for which the full 
density matrix takes the form
\ba
\rho(t) &=& \rho^{(0)}(t)+ \rhoOne(t) =\epsilon(t) |0\rangle \langle 0| + \rhoOne(t).  
\label{rhoExample}
\ea
The single-particle component of the density matrix \rhoOne can be characterized using density matrix components $\rho_{AB}(x,y,t)$
defined as $\rho_{AB}(x,y,t)=${Tr}$[\rhoOne(t) \psiOp_A\dr(x) \psiOp_B(y)]$, i.e., 
\ba
\rhoOne(t) = \sum_{\rs  AB} \integral{x} \integral{ y }
 \rho_{\rs AB}(x,y,t) \psiOp_{\rs B}^\dagger(y) |0\rangle \langle 0| \psiOp_{\rs A}(x),
 \label{Xrho1}
\ea
where $\text {AB}\in\{\ESt\ESt,\ESt \pSt,\pSt\ESt,\ESt \rSt,\rSt\ESt,\rSt\pSt,\pSt\rSt,\rSt\rSt\}$. 

In the following, we will only be interested in the evolution of $\rho^{(N)}$.
First, we will show that in such a situation, the description of the system can be substantially simplified in the case of decoherence due to the decay of the excited states. 
Afterwards, we comment on the impact of dephasing on the system evolution.

\subsection{Decay}
Here, we consider decoherence in the system due to the finite lifetime of the excited states.
For the sake of simplicity, we analyze the decay on the example of the intermediate $\pSt$-state with the decay rate $2\gamP$.
The Louvillian for such a process reads
\ba
\mathcal{L}_{\pSt g}=-\gamP
\integral{y}
\left[
\psiOpP\dr(y)\psiOpP(y)\rhoOp+\rhoOp\psiOpP\dr(y)\psiOpP(y)-2\psiOpP(y)\rhoOp\psiOpP\dr(y)
\right]
\label{Lpg}.
\ea
The first two terms describe the decay of the probability that the system contains $N$ excitations.
The last term depicts the ``quantum jump'' from the ($N+1$)-excitation manifold to the $N$-excitation manifold. Since we consider the case $\rho^{(N+1)}=0$, this process  can be neglected.
Using this observation we can rewrite Master equation as
\ba
\dot{\rhoOp}^{(N)}=-\frac{i}{\hbar} (\Hnon \rhoOpN-\rhoOpN \Hnon\dr)
\label{masterEff},
\ea
where we defined the non-Hermitian Hamiltonian
\ba
\Hnon=\HCoh-i\hbar\gamP 
\integral{y}{\psiOpP\dr(y)\psiOpP(y)
}.
\ea
Next, we write density operator in the general form 
$ \rhoOpN=\sum_{j}p_j\ketbra{\Phi_j}{\Phi_j} $.
Together with \eqref{masterEff} we see that, rather than solving  the Master equation \eqref{masterEff}, we can  solve the
\Schroedinger equation 
\beq
i\hbar \frac{d}{dt}\ket{\Phi_j}=\Hnon \ket{\Phi_j}
\eeq
for the pure state $\ket{\Phi_j}$, which is much more convenient.
Note that 
there are no approximations in this simplification. The only assumption is that we can neglect the occupation of any Hilbert subspace with more than $N$ excitations and that we are only interested in the time evolution of $\rho^{(N)}$. 
Let us illustrate this simplification on the previously described
 example of a single incoming photon  \eqref{rhoExample}. Assuming that at initial time $t=0$ the excitation can be described by a pure state $|\psi_1(0)\rangle$, the full  density matrix simplifies to
\ba
\rho(t) &=&   \epsilon(t) |0\rangle \langle 0| +|\psi_1(t)\rangle \langle \psi_1(t)| \label{XXeq:rhoP}.
\ea
Note, that due to the non-Hermitian nature of the effective Hamiltonian, the probability leaks from the single excitation subspace. 
It corresponds to an increase in time of the probability $\epsilon(t)$ to have zero excitations.

Analogously to the decay of the $\pSt$-level, we can include the decay $2\gS$ of the Rydberg $s$-state. Together with \eqref{quadraticHamiltonian-1D} leads to the non-Hermitian Hamiltonian of the form
\begin{equation}
 H_{0} = \hbar \integral{z} \left(
     \begin{array}{c}
      	    \psi_{e}\\
	    \psi_{p}\\
	    \psi_{s}
     \end{array}\right)^{\dag}
     \left(\begin{array}{ccc}
   - i  c \partial_{z}& g    &  0 \\
         g   &   \delta -i\gamP&  \Omega\\
        0  &  \Omega & -i\gS
    \end{array} \right)
        \left(
     \begin{array}{c}
      	    \psi_{e}\\
	    \psi_{p}\\
	    \psi_{s}
     \end{array}\right).
     \label{quadraticHamiltonian-1D}
\end{equation}
Note, that even though for typical experimental conditions $\gS\ll\gamP$, it can be the decoherence of the Rydberg level that has a leading impact on the losses of a single 
photon inside the medium at the two-photon resonance. 

\subsection{Dephasing}
{In general there exist processes which decrease coherences of the density matrix without affecting the populations. 
In this paper,  we call such processes \textit{dephasing}}.
In Rydberg-EIT setups the dephasing can result from a variety of sources, for example,  finite linewidth of the laser field,  atom-atom interactions or motion of the atoms.
In general, dephasing can not be rigorously treated  by an imaginary part in an effective Hamiltonian. 
In this section we show the impact of dephasing on the description of polaritons propagation.

We start with the Liouvillian describing the dephasing \cite{Fleischhauer2005a,ClaudeCohen-Tannoudji2004} of the \rSt-state
\ba
\mathcal{L}_{ss}=-\gSS\sum_j \left(
\sSS_{j}\sSS_{j}\rhoOp+\rhoOp\sSS_{j}\sSS_{j}-2\sSS_{j}\rhoOp\sSS_{j} 
\right)
\ea
with  $\sSS^{j} = \ketbra{s}{s}_j$ being the projection onto the Rydberg state.
In second quantization, and written using field operators $\psiOpS$, it takes the form
\ba
\mathcal{L}_{ss}=-\gSS
\integral{y}
\left(
{\psi_{s}\dr(y)\psi_{s}(y)\rhoOp}{}
-
\frac{\psi_{s}\dr(y)\psi_{s}(y)\,\rhoOp\,\psi_{s}\dr(y)\psi_{s}(y) }{n(y)}+h.c.
\right).\nn
\ea
In general, the second term in the parentheses is  nonzero even for $\rho^{(N)}$. 
To better understand  the impact of this term, we analyze an exemplary time evolution of a single excitation $\rhoOne$.
For this purpose, we use the representation of the density matrix given by \eqref{Xrho1}.
Next, we project the Master equation onto different components $\rho_{\aInd\b}$ of the single excitation subspace:
\ba
\dot{\rho}_{\aInd\b}(x,y,t)= \braketop{\psi_\aInd(x)}{\dot{\rho}}{\psi_\b(y)}=
\braketop{\psi_\aInd(x)}{
-\frac{i}{\hbar}[\HCoh,\rho]+\mathcal{L}_{ss}(\rho)+\mathcal{L}_{sg}(\rho)
}{\psi_\b(y)},
\label{masterProjected}
\ea
where $\ket{\psi_\b(y)}={\psi_\b\dr(y)}\ket{0}$. We also included the decay of the \rSt-state by the Liouvillian $\L_{sg}$ analogous to $\L_{pg}$  for the decay of the \pSt-state, see Eq. \eqref{Lpg}.

Specifically, the equation of main interest, i.e., for $\rhoSS$ takes the form
\ba
\partial_t \rhoSS(z,z',t)&=&i\O 
\left(\rhoES(z,z',t) + \rhoSE(z,z',t)\right)-\gS \rhoSS(z,z',t)\\ \nn 
&&-
\gSS \rhoSS(z,z',t)+\gSS \rhoSS(z,z',t)\frac{\d(z-z')}{n(z)}.
\ea
Where the last term in the first line depicts the decay due to the finite lifetime, while the  two  terms in the second line describe dephasing.
We see that the dephasing differs from the decay by the last term which is nonzero for $z=z'$. 
Because of this difference one can not use an effective non-Hermitian Hamiltonian for the rigorous description of dephasing, and we will not consider dephasing  in the rest of the manuscript.

\section{Interaction between polaritons}
In this section, we include the strong interaction between the Rydberg atoms. 
It takes the form
\begin{eqnarray}
 H_{\rs rr} & = & \frac{1}{2} \sum_{i\neq j} \VthreeD({\bf x}_{i}- {\bf x}_{j}) P_{i} P_{j}\\
  & = & \frac{1}{2} \integral{{\bf x}} \integral{{\bf y}} \VthreeD({\bf x}-{\bf y}) :  \psiCond_{r}^{\dag}({\bf x}) \psiCond_{r}({\bf x})  \psiCond_{r}^{\dag}({\bf y}) \psiCond_{r}({\bf y}): \nonumber
\end{eqnarray}
with $P_{i} = |\rSt \rangle\langle \rSt|_{i}$ the projection onto the Rydberg state. The notation $\: : \hspace{8pt} :\:$ means normal ordering, which is included
in order to avoid self-interactions.  On the same level of approximation as for the non-interacting Hamiltonian $ H_{0}$, this interaction can then be expressed in terms of the bosonic fields
as 
\begin{equation}
H_{\rs rr} = \frac{1}{2}\integral{{\bf x}} \integral{{\bf y}} \VthreeD({\bf x} - {\bf y}) : \psiOpS^{\dag}({\bf x}) \psiOpS({\bf x}) \psiOpS^{\dag}({\bf y}) \psiOpS({\bf y}):
\label{interaction}
\end{equation}
with the bosonic field operator $\psiOpS^{\dag}({\bf x})$ creating an \rSt-excitation at position $\bx$.

In the one-dimensional limit, the interaction between the Rydberg levels is described by
\begin{equation}
   H_{\rs rr} = \frac{1}{2} \integral{z} \integral{z}' V(z-z')  \psi_{s}^{\dag}(z) \psi_{s}^{\dag}(z') \psi_{s}(z') \psi_{s}(z),
   \label{oneDInteraction}
\end{equation}
where the interaction potential $V$ results from the microscopic interaction potential by an average over the transverse modes
\begin{equation}
   V(z) = \integral{\bR}  \integral{\bR}' \frac{n({\bf R}) n({\bf R}')}{\bar{n}^2}  |u_{\perp}({\bf R})|^2  |u_{\perp}({\bf R}')|^2\VthreeD( {\bf R}-{\bf R}',z) .
\end{equation}
Note, that the transverse mode spacing naturally introduces a cut-off to the van der Waals interaction. 
In this manuscript, we neglect this effect by taking  $V(z) = C_{6}/z^6$, which is an excellent approximation for high Rydberg states, such that the blockade radius is greater than the size of the transverse mode. For additional insights see Ref. \cite{Moos2015a}. 

In the following, the quadratic Hamiltonian
in Eq.~(\ref{quadraticHamiltonian-1D}) for the bosonic fields $\psiOpE$, $\psiOpP$, and $\psiOpS$ together with the interaction $H_{\rs rr}$ in Eq.~(\ref{oneDInteraction})
allows us to apply standard diagrammatic Green's function techniques to study the
properties of the system. The only relevant approximations are that the light modes are restricted to low energies 
$\hbar\omega\ll \hbar \omega_{c}$, and that the photonic density is always much smaller than the atomic density $n({\bf x})$.

\subsection{Equations of motion in \Schroedinger picture}
Here, we derive the equations of motion within the \Schroedinger picture using an effective non-Hermitian Hamiltonian.
The \Schroedinger equation has the form
\ba
i\hbar\partial_t\ket{\psi(t)}=(\Hnon+ H_{\rs rr} )\ket{\psi(t)}.
\label{Schroedinger}
\ea
As an exemplary case, we present the analysis for the wavefunction describing two excitations,
\ba 
\ket{\psi(t)}&=& \integral{x}\integral{y} {\bigg [} \left. \psiWf_{ e\pSt}(x,y,t)  \psiOp_{e}^\dagger(x)   \psiOp\dr_{\pSt}(y)+
 \psiWf_{ e\rSt}(x,y,t)  \psiOp_{e}^\dagger(x)   \psiOp\dr_{\rSt}(y)\right.\nn\\ 
&+&\left.   \psiWf_{\pSt\rSt}(x,y,t)  \psiOp_{\pSt}^\dagger(x)   \psiOp\dr_{\rSt}(y)+
\frac{1}{2}     \psiWf_{ee}(x,y,t)  \psiOp_{e}^\dagger(x)   \psiOp\dr_{e}(y)+\right.\nn\\
&+&\left.  \frac{1}{2}    \psiWf_{\pSt\pSt}(x,y,t)  \psiOp_{\pSt}^\dagger(x)   \psiOp\dr_{\pSt}(y)+
\frac{1}{2}        \psiWf_{\rSt\rSt}(x,y,t)  \psiOp_{\rSt}^\dagger(x)   \psiOp\dr_{\rSt}(y) \right. {\bigg]}
        \ket{0}
. \ea
We will arrive at equations of motion for the two particle amplitudes
by projecting Eq. \eqref{Schroedinger} onto all possible components $\ket{\psiOp_{\rs A}(z)   \psiOp_{\rs B}(z')}=\psiOp\dr_{\rs A}(z)   \psiOp\dr_{\rs B}(z')\ket{0}$. 
For example, the time evolution of 
 $\psiWf_{ep}(z,z')$ is given by
 $\bra{\psiOp_{ e}(z)   \psiOp_{ p}(z')}(-i\hbar\partial_t+H_0+H_{\rs rr})\ket{\psi(t)}=0$.
Without loss of generality, we take $\psiWf_{ ee}(x,y)=\psiWf_{ ee}(y,x)$, $\psiWf_{ \pSt\pSt}(x,y)=\psiWf_{ \pSt\pSt}(y,x)$ and $\psiWf_{ \rSt\rSt}(x,y)=\psiWf_{ \rSt\rSt}(y,x)$.
  The full set of equations has the form
 \ba
i \partial_t \psiWf_{ee}(z,z') &=& -ic(\partial_z+\partial_{z'})\psiWf_{ee}(z,z')	+	g(\psiWf_{ep}(z,z')+\psiWf_{ep}(z',z)),\nn\\
i \partial_t \psiWf_{e\pSt}(z,z')&=& (-ic\partial_z+\D)\psiWf_{e\pSt}(z,z')	+	g(\psiWf_{ee}(z,z')+\psiWf_{pp}(z,z'))+\Omega\psiWf_{es}(z,z'),\nn\\
i  \partial_t \psiWf_{e\rSt}(z,z')&=& (-ic\partial_z-i\gS)\psiWf_{e\rSt}(z,z')	+	g\psiWf_{\pSt\rSt}(z,z')+\Omega\psiWf_{ep}(z,z'),\nn\\
i  \partial_t \psiWf_{\pSt\pSt}(z,z')&=& 2\D\psiWf_{\pSt\pSt}(z,z')	+	g(\psiWf_{e\pSt}(z,z')+\psiWf_{e\pSt }(z',z))+\Omega(\psiWf_{\pSt\rSt}(z,z')+\psiWf_{\pSt\rSt}(z',z)),\nn\\
i  \partial_t \psiWf_{\pSt\rSt}(z,z')&=& (\D-i\gS)\psiWf_{\pSt\rSt}(z,z')	+	g\psiWf_{e\rSt}(z,z')+\Omega(\psiWf_{\pSt\pSt}(z,z')+\psiWf_{\rSt\rSt}(z,z')),\nn\\
i  \partial_t \psiWf_{\rSt\rSt}(z,z')&=& -i2\gS\psiWf_{\rSt\rSt}(z,z')	+\Omega(\psiWf_{\pSt\rSt}(z,z')+\psiWf_{\pSt\rSt}(z',z))+V(z-z')\psiWf_{\rSt\rSt}(z,z').\nn
 \ea

 Alternatively, one can describe the system in the Heisenberg picture. 
 Then, the equation of motion for the time dependent field operators $\psiOp_A(z,t)$ can be derived from Heisenberg-Langevin equations
 \ba
 \partial_t \psiOp_e &=& \frac{i}{\hbar}[\HCoh+\HInt,\psiOp_e],\\
 \partial_t \psiOp_\pSt&=& \frac{i}{\hbar}[\HCoh+\HInt,\psiOpP]-\gamP \psiOp_\pSt+\FOp_\pSt,\\
  \partial_t \psiOp_\rSt&=& \frac{i}{\hbar}[\HCoh+\HInt,\psiOpS]-\gS \psiOp_\rSt+\FOp_\rSt,
 \ea
 where $\HCoh+\HInt$ describes the coherent evolution, see Eqs \eqref{quadraticHamiltonianCoh} and \eqref{oneDInteraction}, while $\FOp_\pSt$ and $\FOp_\rSt$  are the Langevin noise operators corresponding to the decay rates $\gamP$ and $\gS$, respectively.
These equations are the starting point of the analysis presented, for example, in Refs~\cite{Gorshkov2011a,Peyronel2012,Otterbach2013,Moos2015a}.

\section{Diagrammatic methods}
The microscopic Hamiltonian describes three bosonic fields with a non-interacting quadratic part [Eq. \eqref{quadraticHamiltonian-1D}] and a quartic interaction [Eq. \eqref{oneDInteraction}]. 
In the past, such systems have been studied extensively using diagrammatic methods; see for example \cite{Abrikosov:QFT}. 
However, it is important to stress that  the quadratic
Hamiltonian exhibits a rather unconventional form, as the only dynamics is given by the light velocity of the photon. It is this property, which is crucial for the following analysis using diagrammatic methods and gives rise to novel phenomena.

In our previous work \cite{Bienias2014}, we successfully applied diagrammatic methods to the case of two copropagating polaritons. 
Here, we first use the diagrammatic formalism 
to describe
 a single polariton propagating in an external potential.
We show that this problem can be solved by an exact summation of all Feynman diagrams.
Then, we will show an analytical solution of the two-body problem in the weakly interacting regime.
This solution will shed light on the losses of dark state polaritons due to the resonant scattering to bright polaritons, which was studied in \cite{Bienias2014}.

\subsection{Dispersion relation}
\begin{figure}[htp]
\includegraphics[width= 1\columnwidth]{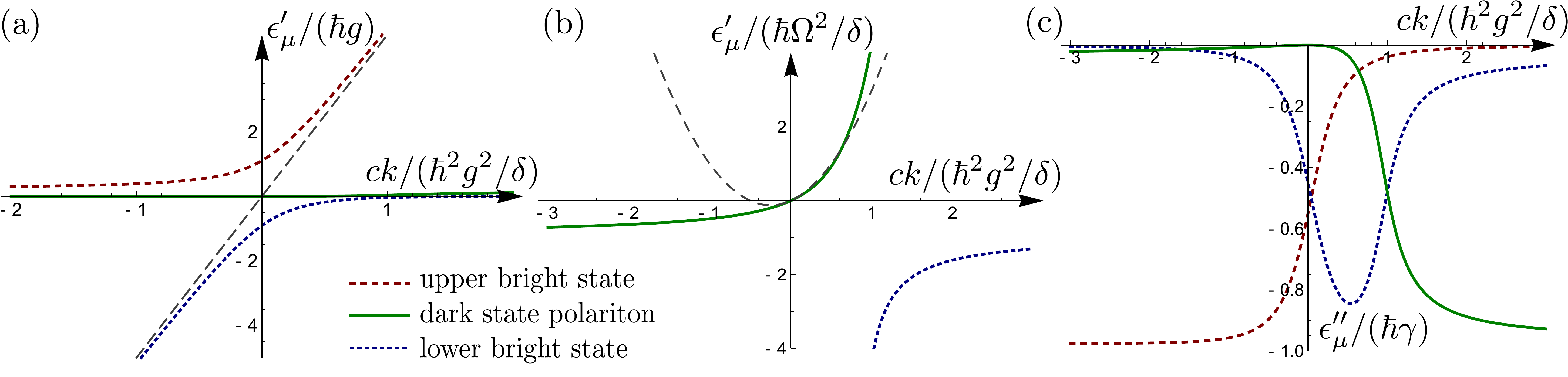}
\caption{Dispersion relation for the three non-interacting polariton branches
for $g =4 \delta$,  $\Omega = 0.25\delta$ and $\Delta =(4-i)\gamP$. (a) The real part of the energy $\epsilon'_\alphaInd(k)$. The gray dashed line depicts light mode dispersion relation. (b) The low energy range of the dispersion relation illustrating dark state polariton behavior: For low momenta the dispersion relation $\epsilon_0(k)$ can be approximated by linear and quadratic contributions.
The gray dashed line shows the contribution from these two terms, see Eq. \eqref{lowMomentaDisp}.
(c) The imaginary part of the energy $\epsilon''_\aInd(k)$: For low momenta the imaginary part of the dark state energy vanishes.
Note different characteristic energy scales for each figure.
}
\label{figDisp}
\end{figure}

First, we analyze the unconventional form of the quadratic Hamiltonian, by looking at its 
spectrum, see Fig~\ref{figDisp}. 
It  is obtained by diagonalizing the quadratic Hamiltonian~(\ref{quadraticHamiltonian-1D}), which reduces to
$H_0= \sum_{\alphaInd \in {0, \pm 1}} \epsilon_{\alphaInd}(\q) \tilde{\psi}^{\dag}_{\alphaInd}(\q) \tilde{\psi}_{\alphaInd}(\q)$.
Here, $\alphaInd \in { \pm1 }$ accounts for the two bright polariton states, while $\alphaInd = 0$ denotes the dark state polariton mode.
The new field operators take the form $ \tilde{\psi}_{\alphaInd \q} = \sum_{\beta \in \{e,p,s\}} U_{\alphaInd}^{\beta}(\q) \psi_{\beta \q} $ with $\alphaInd \in \{0, \pm 1\} $. 
Subsequently, the inverse of $U$, i.e., $\bar{U} \equiv U^{-1}$ provides creation operators $\tilde{\psi}^{\dag}_{\alphaInd \q} = \sum_{\beta \in \{e,p,s\}} \bar{U}_{\beta}^{\alphaInd}(\q) \psi^{\dag}_{\beta \q} $.
Note, that the diagonalizing matrix $U$ is not unitary, due to the imaginary part in the Hamiltonian~(\ref{quadraticHamiltonian-1D}). 
For the clarity of the expressions, we set the decay of the \rSt-state to zero in the rest of the manuscript, i.e., $\g_\rSt=0$. This approximation is well justified for highly excited Rydberg states used in nowadays experiments, because the propagation time of the photon in the medium is much shorter than the life-time of the Rydberg state.

In the regime of {low-momentum and low-energy}, i.e., $\om, v_g k\ll \min[\O,|\D|,g$], the dispersion relation for the dark state polariton is well accounted for by the two terms in
\ba
\epsilon_{0}(\q)= \hbar v_{g} \q - \frac{\hbar^2}{2 m} \q^2,
\label{lowMomentaDisp}
\ea
 with  group velocity and polariton mass
\ba
v_g=\frac{ \Omega ^2}{g^2+\Omega ^2}c, \qquad
m=\frac{\left(g^2+\Omega ^2\right)^3}{2c^2g^2 \Delta  \Omega ^2}
\label{vgAndMass}
.
\ea
Finally, it is worth pointing out that while the general expressions for the dispersion relation are complicated, the expression for momentum as a function of energy has a simple analytical  form 
\beqa
\hbar ck(\omega)
&=&\hbar \omega\left(\frac{g^2}{\O^2}\frac{1}{1+\frac{\D\om-\om^2}{\O^2}}+1\right).
\label{inverseDispersion}
\eeqa

\subsection{Polariton propagation in external potential}
In this section, building on the understanding of  single-body physics, we describe the polariton propagation in an external potential $V(z)$, acting only on the Rydberg \rSt-state
\ba
H_{\rs ext} = \integral{z}\psiOpS(z)\dr V(z) \psiOpS(z).
\ea
This potential can be a result of an interaction  between the polariton and  a stationary Rydberg excitation in state $\ket{\rSt'}$. 
Such a configuration is relevant for recent experimental realizations of single photon switch and transistor \cite{Baur2014,Tiarks2014,Gorniaczyk2014,Gorniaczyk2015}.
Note, that in order to neglect readout of the stored excitation, the state $\ket{\rSt'}$ has to be  different than the state $\ket{\rSt}$ in the EIT scheme.
Moreover, alternative ways of treating this problem can be found, for example, in Ref. \cite{Gorshkov2011a,Li2014a}.

We start by pointing out, that
the Hamiltonian  conserves the total energy $\hbar \omega $.
Then, the single polariton scattering properties  can be well accounted for using the $T$-matrix formalism.
As the interaction acts only between the
Rydberg states, it is sufficient to study the $T$-matrix for the Rydberg states alone,  which will be denoted as $T_{kk'}(\omega)$.
Here, $\hbar k$ denotes the momentum of the incoming particle, and $\hbar k'$ the momentum of the 
outgoing state. The relation between the $T$-matrix and the \rSt-state amplitude
$\psiWf_s$ by definition
is provided by  the relation
\begin{equation}
\psiWf_{s}(z) = \frac{1}{V(\z)}\integralf{ k'}{2\pi}  e^{i k'z} \: T_{kk'}.
\end{equation}
\begin{figure}[ht]
\includegraphics[width= 0.8\columnwidth]{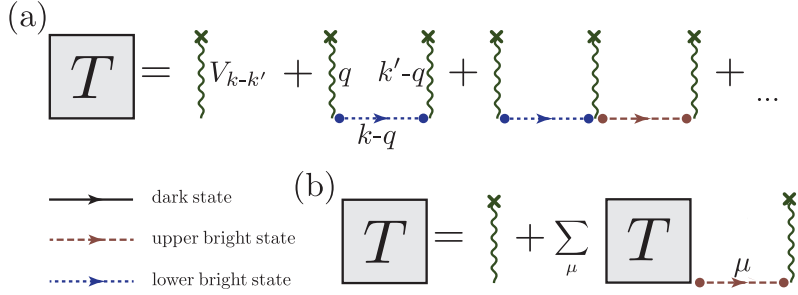}
\caption{(a) Illustration of  ladder diagrams up to the third order: the interaction $V$ is denoted by a wavy line, while the straight lines with an arrow are  Green's functions for the three polariton modes $1/(\hbar \omega - \epsilon_{\aInd} + i \eta)$, and the dots mark the overlap factors $U^{s}_{\aInd}$  and $\bar{U}_{s}^{\aInd}$ of the polariton with the Rydberg state.  The $T$-matrix includes all diagrams up to arbitrary order with all possible intermediate polaritons. (b) Illustration of the integral equation \eqref{tMatrixEq}.
}
\label{figLadder1} 
\end{figure}
For single polaritons, the $T$-matrix is 
expressed as a resummation of all ladder diagrams, \figref{figLadder1},
which gives rise to the integral equation  \cite{Abrikosov:QFT}
\ba
T_{kk'}(\omega) = V(k-k') + \integralf{q}{2\pi} T_{kq}(\omega)   \chi_{q}(\omega) V(q-k').\label{tMatrixEq}
\ea
Here, $\chi_{q}$ denotes the full  propagation of a single polariton and its overlap with the Rydberg state
\begin{equation}
  \chi_{q}\left(\omega\right) = \sum_{\alphaInd\in \{0,\pm 1\}}  \frac{\bar{U}^{\alphaInd}_{s }(q)  U_{\alphaInd}^{s}(q) }{\hbar \omega - \epsilon_{\alphaInd}(q) + i \eta}
  \label{chiQ}
\end{equation}
It is a special property of the polariton Hamiltonian, that 
$\chi_q$
reduces to two terms
\begin{equation}
\chi_q = \bar{\chi}(\om)+ \frac{\alpha(\om)}{\hbar c\, k(\om)-\hbar c\,q+i \eta} 
 \label{chiexpansionA0}.
\end{equation}
Here, $\bar{\chi}(\omega)$ accounts for the saturation of the polariton propagation at large momenta $q\rightarrow \pm \infty$
and takes the  form
\ba
\hbar\bar{\chi}(\omega) &=&\frac{\D}{\O^2}\frac{1-\frac{\omega }{\D}}{1+\frac{\Delta  \omega }{\O^2}-\frac{\omega ^2}{\O^2}},
\label{chiA}
\ea
which for $\om\ll\O^2/'\D'$ simplifies to $\hbar\bar{\chi}(\omega)=\D/\O^2$. Note, that $\D=\d-\i\gamma$ is complex, and takes into account the decay of the intermediate level.
The second term in Eq.~(\ref{chiexpansionA0}) characterizes  the pole structure of the propagating 
polariton. This term reduces to the propagator of a single polariton with momentum $\hbar k(\om)$, given by \eqref{inverseDispersion}, and $\alpha$ 
depends on the energy $\omega$ of the incoming polariton:
\beqa
\a(\omega)
&=&\frac{g^2}{\O^2 }\frac{ 1}{\left((\Delta -\omega ) \omega/\O^2 +1\right)^2}.
\eeqa
In order to eliminate the saturation-term $\bar{\chi}$, we Fourier transform the \TM-matrix equation \eqref{tMatrixEq} to real space
\beqa
\TM_{k}(\z)&=&V(x)e^{ik\z}+\chiB(\om) V(\z)\TM_{k}(\z) +V(\z)\integral{y}G(\omega,{\z-y})
\TM_{k}(y)\nn
\eeqa
with ${G}(\om,\z)=-i\alpha(\om)\,\theta(\z)e^{ik(\om)\z}$ being the Fourier transform of the second term in Eq. \eqref{chiexpansionA0}, where $\theta(\z)$ is the Heaviside step function.
Introducing the effective interaction potential 
\begin{equation}
\Veff (\z) = \frac{V(\z)}{1- \bar{\chi}(\omega) V(\z)} ,
  \label{effectiveinteraction}
\end{equation}
the equation for the $T$-matrix reduces to
\beqa
\TM_{k}(\z)&=&\Veff(\z)e^{ik\z}\left(1-i\aHbarC\integralb{-\infty}{\z}{y}e^{-iky}
\TM_{k}(y)\right).
\eeqa
This equation can be solved analytically leading to the expression for \TM-matrix
\beqa
\TM_{k}(\z)&=&e^{ik\z}\Veff(\z)\exp\left[ {-i\aHbarC\integralb{-\infty}{\z}{y} \Veff(y)} \right].
\label{eq:gammaKX} 
\eeqa

\begin{figure}[htp]
\includegraphics[width= \textwidth]{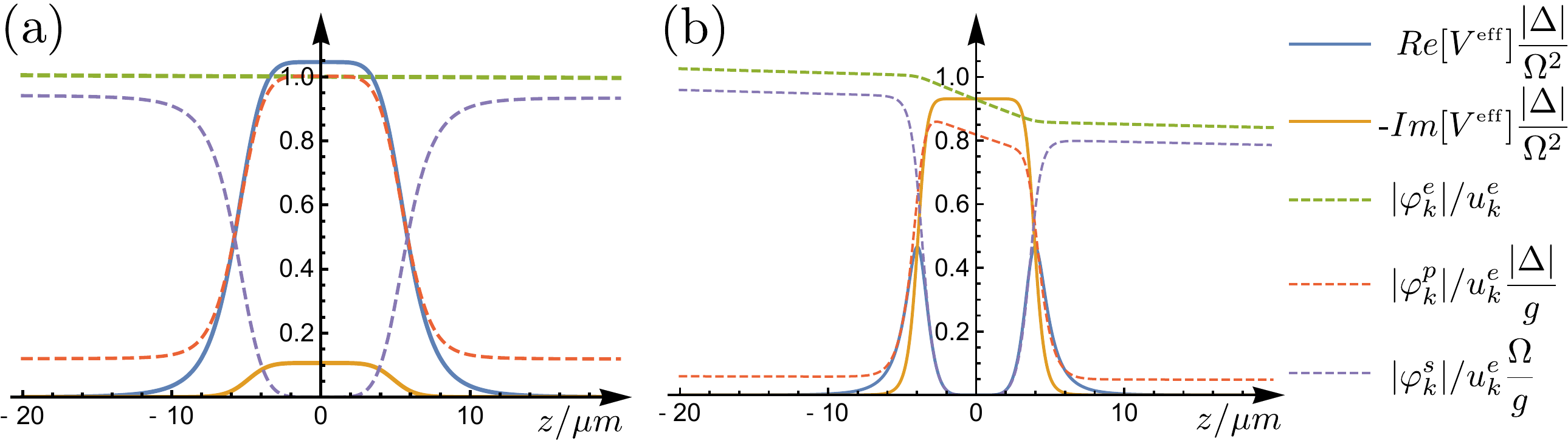}
\caption{
Wavefunction components and effective interaction in function of distance. (a) The dispersive regime with $\D=(10-i)\g$, (b) The dissipative regime with $\D=-i\gamma$.
All other parameters are the same for both regimes:  $\om=0.05\O^2/|\D|$, $g=1500 \gamP$, $\O=1.4 \gamP$, $C_6=3.3\times10^4\gamP\; \mu m^6$.
}
\label{components}
\end{figure}

Based on the solution for $T_k(\om)$ we can derive all components of the wavefunction describing a single polariton.
For this purpose, we start from the relation between the \TM-matrix and the outgoing state
\beqa
\psiWf_k^\b(\z)=
e^{ik\z}{u}_{k}^\b+u_k^s\integralf{q}{2\pi}e^{iq\z}\TM_{kq}
\chiBetaSq
,
\label{eq:PsiKX2}
\eeqa
where the index $\b\in\{e,p,s\}$ depicts components of the incoming $u_k^\b$ and the outgoing $\psiWf_k^\b$ states. 
In order to arrive at formula \eqref{eq:PsiKX2} we used the fact that the only non-vanishing  element of the \TM-matrix is between $s$-states. 
Moreover, we introduced $\chiBetaSq$ which is the generalization of $\chi_q(\omega)$, see Eq. \eqref{chiQ}, and describes the propagation of a single polariton and its overlap with $s$-state and \b-state 
\ba
\chiBetaSq= 
 \sum_{\alphaInd\in \{0,\pm 1\}}  \frac{\bar{U}^{\alphaInd}_{s }(q)  U_{\alphaInd}^{\beta}(q) }{\hbar \omega - \epsilon_{\alphaInd}(q) + i \eta}.
\ea
Moreover, analogously to Eq. \eqref{chiexpansionA0}, \chiBetaSq\, can be re-written in the following form
\begin{equation}
\chiBetaSq=  \bar{\chi}^\b(\om)+ \frac{\alpha^\b(\om)}{\hbar ck(\om)-\hbar cq+i \eta} 
 \label{chiexpansionB}.
\end{equation}
Note that in the newly introduced notation, by definition, the following relations are satisfied: $\bar{\chi}^s\equiv\bar{\chi}$ and $\a^s\equiv\a$.
Next, we Fourier transform Eq. \eqref{eq:PsiKX2}, and then insert to it the solution for \TM-matrix, given by Eq. \eqref{eq:gammaKX}.
Furthermore, we use the relation $u_k^s/u_k^\b=\a^s/\a^\b$ and finally arrive at the expressions for the wavefunction components 
\beqa
\psiWf_{k}^{\gamma}(\z)=e^{ik\z}u^{\gamma}_{k}\exp\left[ {-i\aHbarC\integralb{-\infty}{\z}{y} \Veff(y)} \right]\left(1+
\Veff(\z)\chiB^{\gamma}\frac{\alpha}{\alpha^{\gamma}}
\right).
\label{eq:PsiKX35}
\eeqa
From this solution, we see that for distances much larger than the range of the interaction the outgoing state is proportional to the incoming one. Hence, due to the interaction, all components pick up a common exponent.
Next, we comment on the form of each component separately. For van der Waals interaction $V(r)=C_6/r^6$,  all of them are shown in \figref{components}. 

First, for the photonic component the saturation vanishes, i.e. $\chiB^e=0$, which leads to
\beq
\psiWf_{k}^{e}(\z)=e^{ik\z}u^{e}_{k}\exp\left[ {-i\aHbarC
\integralb{-\infty}{\z}{y} V_e(y)} \right].
\eeq
We see that even close to the impurity the photonic component only picks up a phase factor as a result of the interaction with the impurity.
Note, that due to the finite $\gamP$ this phase factor is complex what leads to the decay of $\psiWfE_k$.

Secondly,  for the Rydberg component expressed using $u_k^e$ we arrive at
\beqa
\psiWf_{k}^{s}(\z)=-e^{ik\z}
\frac{g }{\O}
\frac{1 }{1+\frac{\omega  \Delta}{\O^2} -\frac{\omega^2}{\O^2} }
\exp\left[ {-i\aHbarC\integralb{-\infty}{\z}{y} V_e(y)} \right]\frac{1}{1-\chiB^sV(\z)}
u^{e}_{k},
\label{eq:PsiKX4}
\eeqa
from which we see that the Rydberg component is suppressed at distances shorter than the so-called Rydberg blockade $\xi$ defined via $|V(\xi)\bar{\chi}|=1$. 
The reason is the following: At short distances, due to the interaction, the Rydberg-level is shifted out of resonance and can not be excited \cite{Lukin2001}. 

Finally, the $p$-state component has the form
\beqa
\psiWf_{k}^{p}(\z)&=&-e^{ik\z}
\frac{g}{\O}
\frac{1}{1+\frac{\Delta \omega  }{\O^2} -\frac{\omega^2 }{\O^2}}
\frac{1}{\O}\frac{{V}{}-{\omega }{}}{1-\bar{\chi}V}
\exp\left[ {-i\alpha\integralb{-\infty}{\z}{y} V_e(y)} \right]
u^{e}_{k}.
\eeqa
This component vanishes for distances much greater than the blockade length, i.e.,  $x\gg\xi$, as long as $\om\ll\O^2/|\D|$.
The last condition corresponds to the EIT transparency condition. Once this condition is broken, the polariton has a significant admixture of the $p$-state, which causes the decay of the polariton inside the medium. 
Moreover, for short distances $\z < \xi$ with $\om\ll\O^2/|\D|$, the \pSt-state component saturates at  $|\psiWf_{k}^{p}|\sim \frac{g}{|\D|} u_k^e$.
Hence, in the dissipative regime with small detuning $\d<\gamP$ the \pSt-component is larger than in dispersive regime with $\d\gg\gamP$. It corresponds to smaller losses in the dispersive regime, as shown in \figref{components}.

\section{Two body problem}
This section deals with two photons copropagating in the Rydberg-EIT medium, see \figref{fig1}. We first review the general approach to this problem using Feynman diagrams, shown in \cite{Bienias2014}.  Based on this description we, afterwards,  present the analytical solution in the weakly interacting regime.

\begin{figure}[htp]
\includegraphics[width= 0.65\columnwidth]{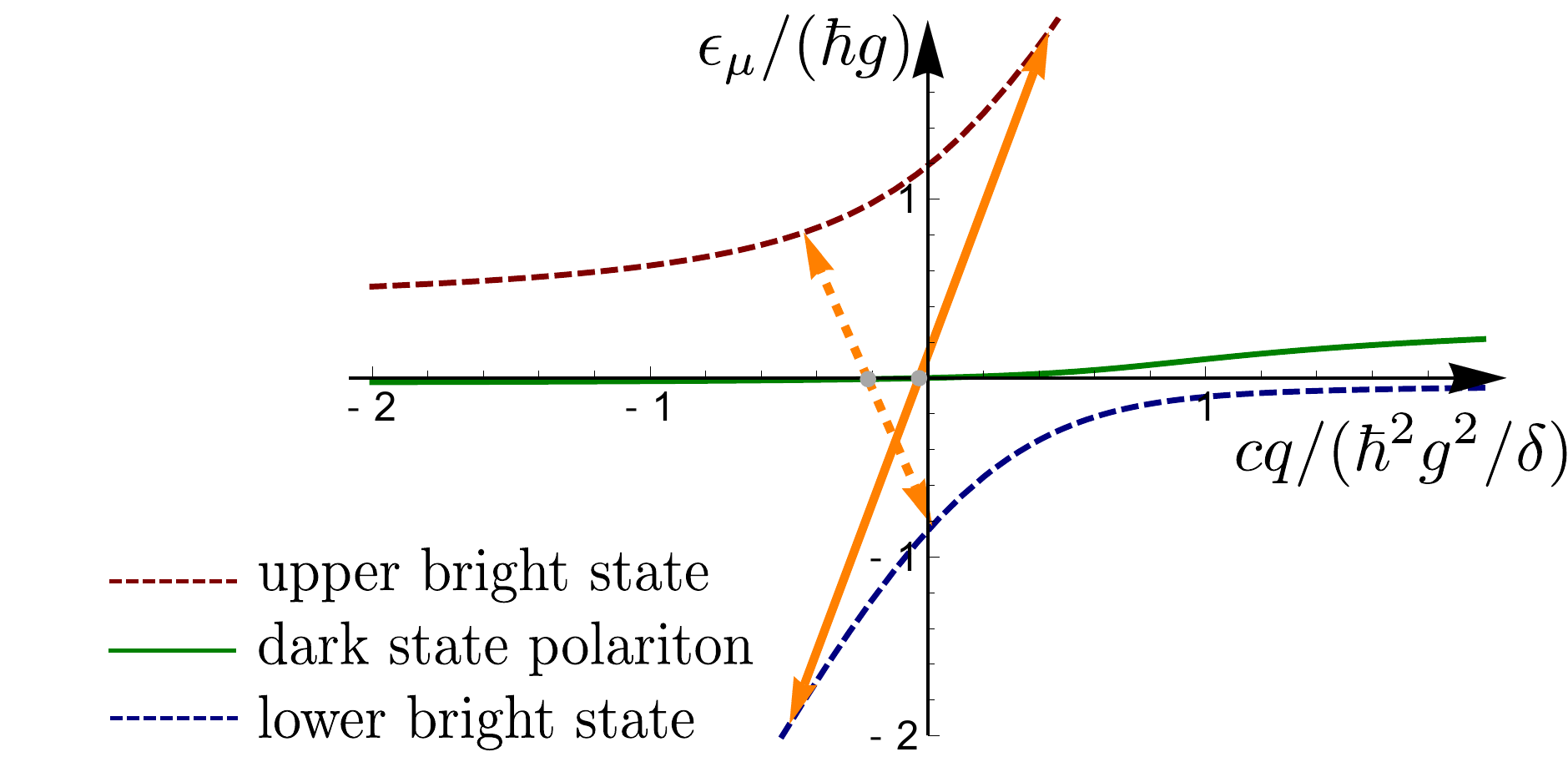}
\caption{ 
Illustration of a resonant excitation of two bright polaritons from two dark polaritons for $g =3 \delta$,  $\Omega=\delta/3$ and $\gamP=0$.
 Two different cases of the total energy $\hbar \omega$ of the incoming dark-state polaritons are shown.
In both situations the relative momentum  is zero, $\hbar k=0$.
Orange solid arrow shows the case of $\om=-0.06\O^2/\d$, while orange dashed line the case of  $\om=-0.35\O^2/\d$.
Resonant excitation conserves center of mass momentum $\hbar K$, as well as the total energy, $\hbar \omega=2\epsilon_0(K/2)=\epsilon_-(K/2-\kB/2)+\epsilon_+(K/2+\kB/2)$, where $\hbar \kB = \pm \sqrt{\hbar  \bar{\omega}_{\rs B}m}$. 
}
\label{fig6}
\end{figure}

The two-polariton scattering properties  are well accounted for by the $T$-matrix.
As the interaction $V(r)$ acts only between the
two Rydberg states, it is sufficient to study the $T$-matrix for the Rydberg states alone, denoted as $T_{k k'}(K,\omega)$.
Importantly, the total energy $\hbar \om$ as well as the center-of-mass momentum $\hbar K$ are conserved.
Moreover, in this section, $\hbar k$ is the relative momentum of the two incoming  polaritons and $\hbar k'$ the relative momentum of the
outgoing polaritons, while $\r$ denotes relative coordinate $r=z-z'$.  
For two polaritons, the $T$-matrix is determined by the integral equation \cite{Abrikosov:QFT}
\begin{equation}
T_{kk'}(K,\omega) = V_{k-k'} + \integralf{ q}{2\pi}  T_{kq}(K,\omega)\:  \chi_{q}(K,\omega) V_{q-k'} ,
\label{2bTmatrix}
\end{equation}
which can easily be understood as a resummation of all ladder diagrams,  see Fig.~3(a) in \cite{Bienias2014}.
The full pair propagator of two polaritons and its overlap with the Rydberg state takes the form
\begin{equation}
  \chi_{q}\left(K,\omega\right) = \sum_{\alphaInd,\beta \in \{0,\pm 1\}}  \frac{\bar{U}^{\alphaInd}_{s }(p)  U_{\alphaInd}^{s}(p)   \bar{U}^{\beta}_{s }(p')U_{\beta}^{s}(p') }{\hbar \omega - \epsilon_{\alphaInd}(p) - \epsilon_{\beta}(p') + i \eta},
\end{equation}
with $p=K/2+q$ and $p' = K/2-q$. Due to the special property of our polariton Hamiltonian
the pair propagation reduces to three terms,
\begin{equation}
\chi_q = \bar{\chi}+ \frac{\alpha}{\hbar \bar{\omega}-\hbar^2 q^2/m+i \eta} + \frac{\alpha_{\rs B}}{\hbar \bar{\omega}_{\rs B} -\hbar^2 q^2/m + i \eta}
 \label{chiexpansion}.
\end{equation}
Where mass $m$ is given by \eqref{vgAndMass},
and $\bar{\chi}(\omega)$ accounts for the saturation of the pair propagation at large momenta $\hbar q\rightarrow \pm \infty$
and reads 
\begin{equation}
\bar{\chi}(\omega) =\frac{1}{\hbar} \frac{ \Delta-\frac{\omega}{2} - \frac{\Omega^2}{\Delta - \omega}}{ \omega\left(\Delta - \frac{\omega}{2}\right) + 2 \Omega^2}.
\label{chi}
\end{equation}
The second term in Eq.~(\ref{chiexpansion}) is  the pole structure for the propagation of the two incoming
polaritons. This term reduces to the propagator of a single massive particle, where  $\alpha$ and $\bar{\omega}$
depend on the center-of-mass momentum $\hbar K$ and total energy $\hbar \omega$. 
The latter defines  the relative 
momentum $\hbar k = \pm \sqrt{\hbar \bar{\omega} m}$ of the incoming scattering states. 
For analytical expressions for $\a$ and $\bar{\omega}$ see Appendix in Ref. \cite{Bienias2014}.
Finally, the last term 
accounts for a second pole, describing the phenomenon of resonant scattering of two incoming polaritons into 
a different outgoing channel, \emph{e.g.}, the conversion of two dark  polaritons into an upper and a lower bright polariton, see \figref{fig6}, 
and therefore denoted by `${\rm \scriptscriptstyle B}$'. 

The influence of the second pole can be measured by the dimensionless parameter
$\zeta(K,\omega)=\sqrt{|(\bar{\omega}  \alpha_{\rs B}^2)/(\bar{\omega}_{\rs B}\alpha^2)|} $. In \cite{Bienias2014} we have shown that 
$\zeta(K, \omega) $ is strongly suppressed in several relevant regimes.
In the next subsection, we will show how the parameter $\zeta(K,\omega)$ relates to the solution of two-body problem in the weakly interacting regime.

\subsection{Exact solution for weak interactions}
The interaction strength can be conveniently quantified by the dimensionless parameter $\xi/\lambdabar$, 
where  $\lambdabar =\sqrt{ | \hbar^2\bar{\chi} / (\alpha m)|}$ is the de Broglie wavelength associated with the depth (or height) $|\a/ \bar{\chi}|$ of the effective potential. 
Here, we present exact solution of the two-body problem for weak interactions, i.e., for  $\xi/\lambdabar \ll 1$,
in which case the interaction potential can be replaced by a $\delta$-function. 
We start by rewriting the equation for the \TM-matrix \eqref{2bTmatrix} using the effective potential $\Veff(\r)=V(\r)/(1-\chiB(\om)V(\r))$ and explicitly including the pole structure,
\ba
T_{kk'}(K,\omega) =\Veff_{k-k'} + \integralf{ q}{2\pi}
T_{kq}(K,\omega)
\left( \frac{\alpha }{\hbar \bar{\omega}-\frac{\hbar^2 q^2}{m}+i \eta} + \frac{\alpha_{\rs B}}{\hbar \bar{\omega}_{\rs B} -\frac{\hbar^2 q^2}{m} + i \eta} \right)
\Veff_{q-k'},\nn
\ea
This equation is equivalent to the Lippmann-Schwinger equation for  $\psiWf$, defined by $\psiWf(r) \Veff(r)=\integral { k'}    e^{i r k' }  T_{k k'}/(2 \pi)$,
\begin{equation}
\psiWf(\r)= \psiWf_0(\r)+ \integral{y} \fullG(\r-y) \;\a\:  \Veff(y) \,\psiWf(y).
\label{lippman}
\end{equation}
Note that, analogously to the single polariton, the wavefunction component $\psiWf_{ss}$ describing two Rydberg excitations, can be expressed using \TM-matrix, i.e., $\psiWf_{ss}(r) V(r)=\integral{ k'}    e^{i r k' }  T_{k k'}/(2 \pi)$.
Moreover, the incoming wave  $\psiWf_0(\r)=e^{ik\r}$, and the propagator $\fullG$ in real space has the form
\begin{equation}
\fullG(\r)= -
\frac{i}{2}\frac{m}{\hbar^2}\left(
\frac{e^{ik|\r|}}{k}+
\frac{\a_{\rs B}}{\a}\frac{e^{i\kB|\r|}}{\kB}
\right)
 \label{gFullSpace},
\end{equation}
with $k=\sqrt{\bar{\om}m/\hbar}$ and $\kB=\sqrt{\bar{\om}_{\rs B}m/\hbar}$.
Then, the  solution of \eqref{lippman} can be found via the re-summation of all orders in Born expansion
\ba
\psiWf(\r)&=& \psiWf_0(\r)+ \integral{y} \fullG(\r-y) \;\a\:  \Veff(y) \,\psiWf_0(y)\nn\\
&+& \integral{y} \integral{y'}\fullG(\r-y) \;\a\:  \Veff(y)\;\fullG(y-y') \;\a\:  \Veff(y') \,\psiWf_0(y')+ ... \label{summation3}
\ea
In the case of weak interactions $\xi\ll\lambdabar$, we can replace the effective interaction by the potential $\vDelta\d(\r)$, where $\vDelta=\integral{r} \a \Veff(\r)$. It enables us to simplify the expression for $\psiWf$ to
\ba
\psiWf(\r)&=&\psiWf_0(\r)+\fullG(\r)\vDelta \left(1+\fullG(0)\vDelta + (\fullG(0)\vDelta )^2 +...\right) \nn\\
&=& \psiWf_0(\r)+\vDelta\; \fullG(\r)\frac{1}{1-\vDelta\; \fullG(0)}\nn\\
&=& e^{ik\r} -
\frac{1}{1+\frac{k}{k_B}\frac{\a_B}{ \a}-2i \frac{ \hbar^2 k }{m\vDelta}}
\left(
e^{ik|\r|}+\frac{\a_B}{\a}\frac{k}{\kB}e^{i\kB|\r|}
\right).
\ea
We see that the dimensionless parameter 
$\zeta(K,\omega)={|(k  \alpha_{\rs B})/(\kB \alpha)|}$
controls the influence of the second pole.
Since the term proportional to $e^{i\kB|\r|}$ accounts for the resonant excitation of an upper and lower bright polariton, 
 this process is strongly suppressed for small  parameter $\zeta(K,\omega)\ll1$.

\section{Conclusions}
In conclusion, in the extended introduction, we presented the current state-of-the-art in the research field of Rydberg slow light polaritons.
We derived a microscopic Hamiltonian describing the propagation of Rydberg slow light polaritons in one dimension. 
We described the decay and the dephasing of polaritons within a Master equation approach 
and commented on conditions when the decay can be described using the \Schroedinger equation with effective non-Hermitian Hamiltonian.
We derived equations of motion in \Schroedinger picture and compared it with the commonly used derivation of the Heisenberg-Langevin equations of motion.
Next, we analyzed the dispersion relation of dark and bright polaritons --- the basis of the diagrammatic description of the strongly interacting  polaritons.
We illustrated this method on two examples: 
First, by summation of all Feynman diagrams we derived the exact solution for a single polariton in an external potential.
Secondly, we exactly solved the two body problem  in a weakly interacting regime. 

\begin{acknowledgement}
The author would like to thank H.P. B\"uchler for the fruitful collaboration, 
whereas A. Gaj, K. Jachymski and D. Peter for proof-reading the manuscript.
Financial support from EU Marie Curie ITN COHERENCE
and the H2020-FETPROACT-2014 Grant No. 640378 (RYSQ) is gratefully acknowledged.

\end{acknowledgement}

\bibliographystyle{/Users/admin/Dropbox/notes/latex/bernd}
\bibliography{/Users/admin/Documents/Bibtex/library}

\end{document}